\documentclass{PoS}

\def	\pt	{p_{T}}
\def  \deta {\Delta\eta}
\def  \s {\sqrt{s_{NN}}}

\title{Prospects for identified leading particle correlation at LHC and in ALICE}

\ShortTitle{Present and future of particle correlations}

\author{\speaker{Levente Molnar}\\
        INFN Sezione di Bari, KFKI-RMKI\\
        E-mail: \email{Levente.Molnar@ba.infn.it}}

\abstract{Azimuthal di-hadron correlations play important role in 
the characterization of the medium created in heavy-ion collisions at RHIC.
Moreover, as a novel phenomenon, strong modification of the away-side 
correlation is observed in Au+Au with respect to p+p collisions.
Below the exclusive jet reconstruction threshold at LHC, leading 
particle correlations will provide access to the regime where hard scatterings
and bulk medium properties can be simultaneously studied.
Leading particle correlations can be extended to very low transverse
momenta via the tracking and particle identification capabilities 
of ALICE, to the coalescence and hydrodynamic domains.
In preparation for the first p+p and Pb+Pb collisions of LHC, 
we present prospects on leading particle correlations 
with identified particles in ALICE.
}

\FullConference{High-pT physics at LHC\\
		 March 23-27 2007\\
		 University of Jyv$\ddot{a}$skyl$\ddot{a}$, Jyv$\ddot{a}$skyl$\ddot{a}$, Finland}

\begin{document}

\section{Introduction}
Awaiting for the startup of the next generation particle and heavy-ion physics experiments
at the Large Hadron Collider (LHC), we review few selected results from the Relativistic Heavy Ion Collider (RHIC), where heavy-ion 
collisions at 200 GeV c.m.s. energy provide the most suitable environment, up to now, to study the 
strongly interacting matter under extreme temperature and energy density.
At RHIC, in central Au+Au collisions strong suppression of large transverse momentum
(un)identified charged hadrons is observed with respect to peripheral Au+Au, p+p or d+Au
collisions~\cite{hptsupSTAR,hptsupPHENIX}. This suppression is referred as jet quenching
and extends to large transverse momenta ($\pt \sim$ 15-20 GeV/c).
Furthermore, in azimuthal di-hadron correlations 
strong modification of the away-side correlation is seen in Au+Au 
with respect to p+p collisions~\cite{AwayDisPhenix,AwayDisStar}, 
followed by similar observations even at SPS 
energies~\cite{AwayDisCer}. 
Extending the azimuthal di-hadron correlations to pseudo-rapidity as well, on the near-side, 
long range pseudo-rapidity correlation appears as shown in Fig.~\ref{figure:ridge}.
In the so called intermediate transverse momentum region (2$ <\pt <6$ GeV/c)
enhancement of baryon production is observed in mid-central, central Au+Au collisions 
with respect to p+p collisions~\cite{PhenixPbarPi,StarPbarPi}.
\begin{figure}[th]
\vspace{0.5cm}
\includegraphics[width=0.48\textwidth]{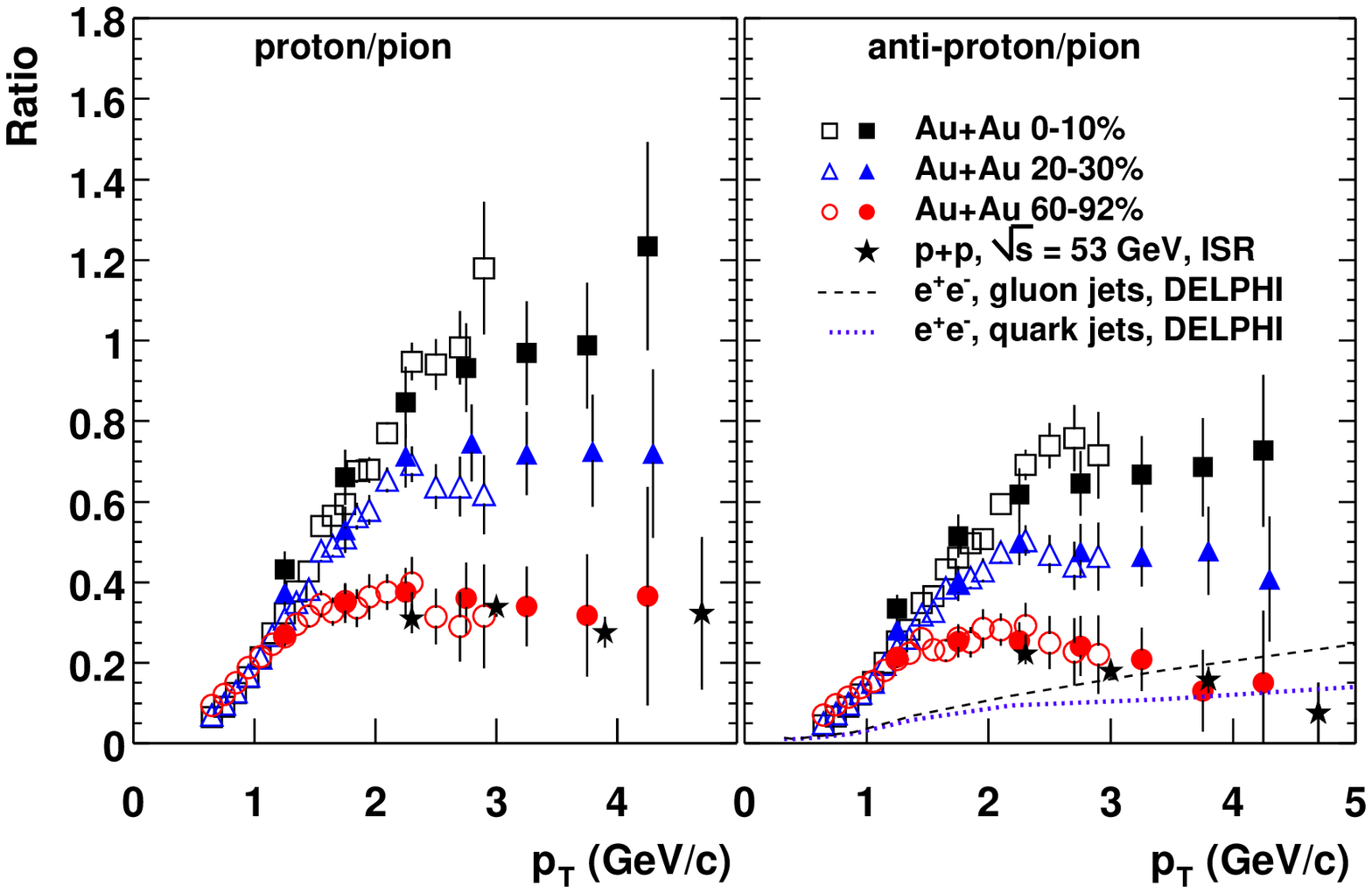}
\includegraphics[width=0.48\textwidth]{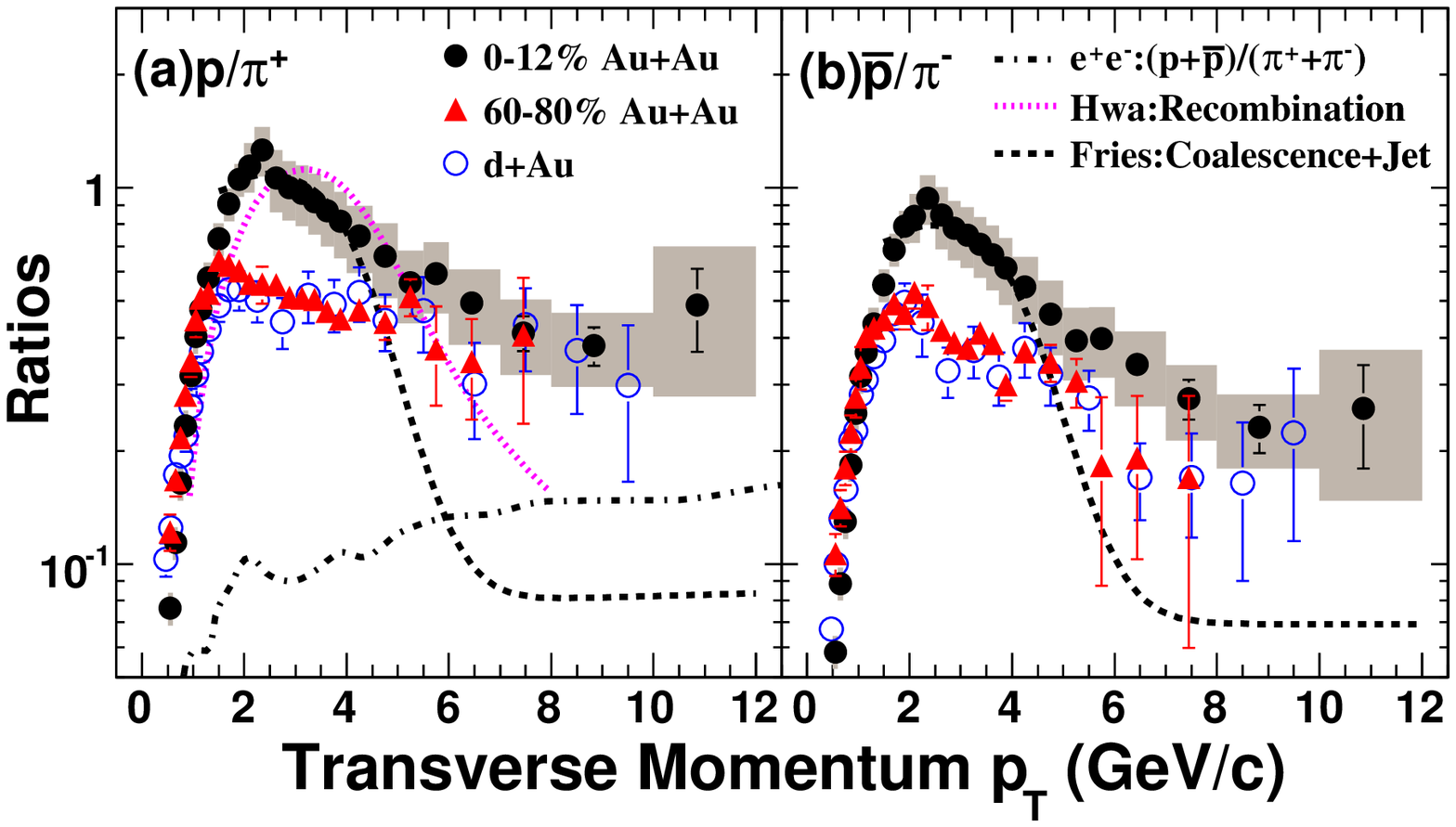}
\caption{p($\overline{p}$)/$\pi^{\pm}$ ratios measured by PHENIX~\cite{PhenixPbarPi} and STAR~\cite{StarPbarPi}.}
\label{figure:pbarpi}
\end{figure}
These measurements highlight the importance of the intermediate $\pt$ region 
where one can investigate to the interaction between the hard probes and the bulk matter created in 
heavy-ion collisions.

\section{Review of RHIC results}
\begin{figure}[th]
\includegraphics[width=0.48\textwidth]{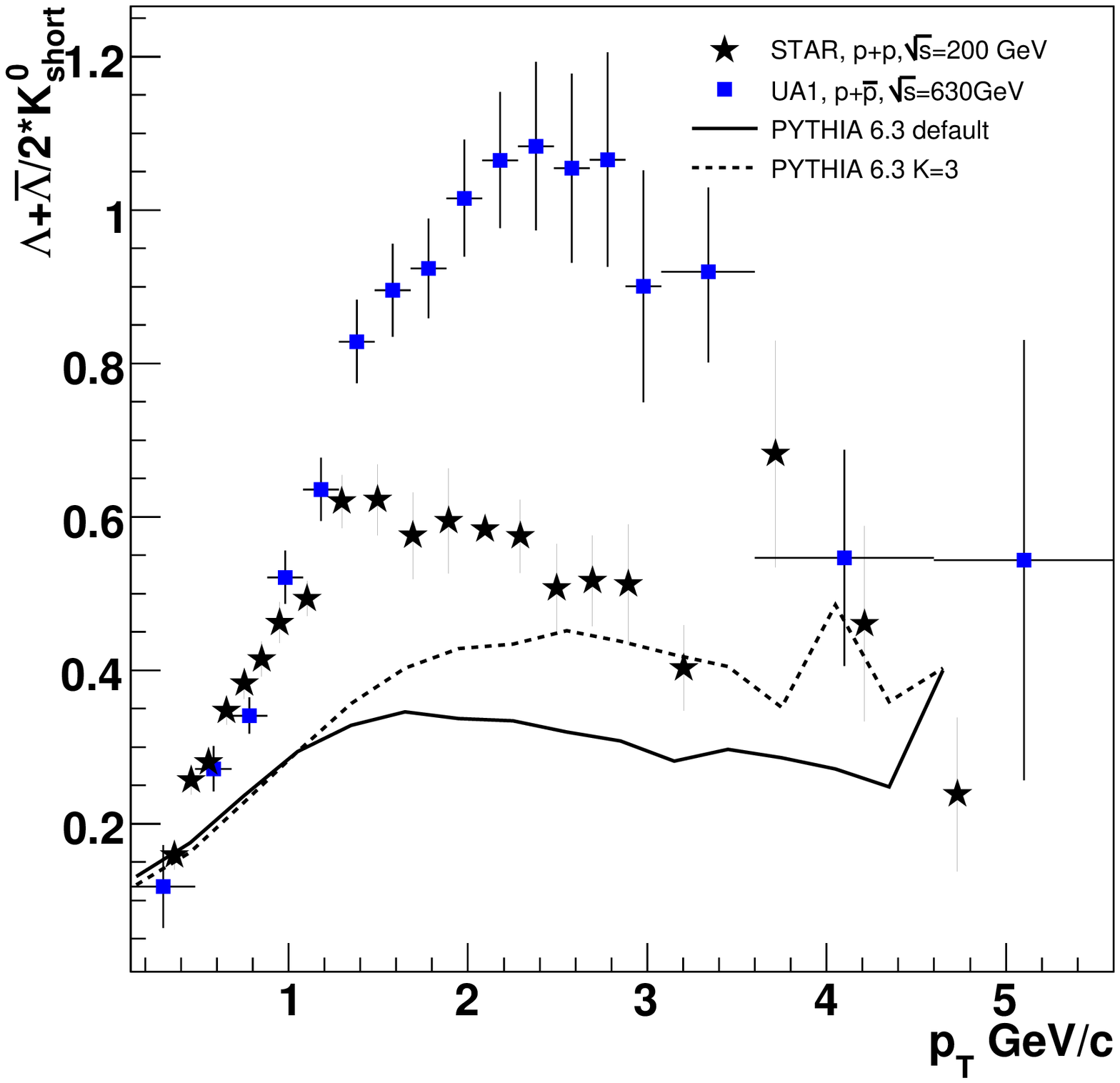}
\includegraphics[width=0.48\textwidth]{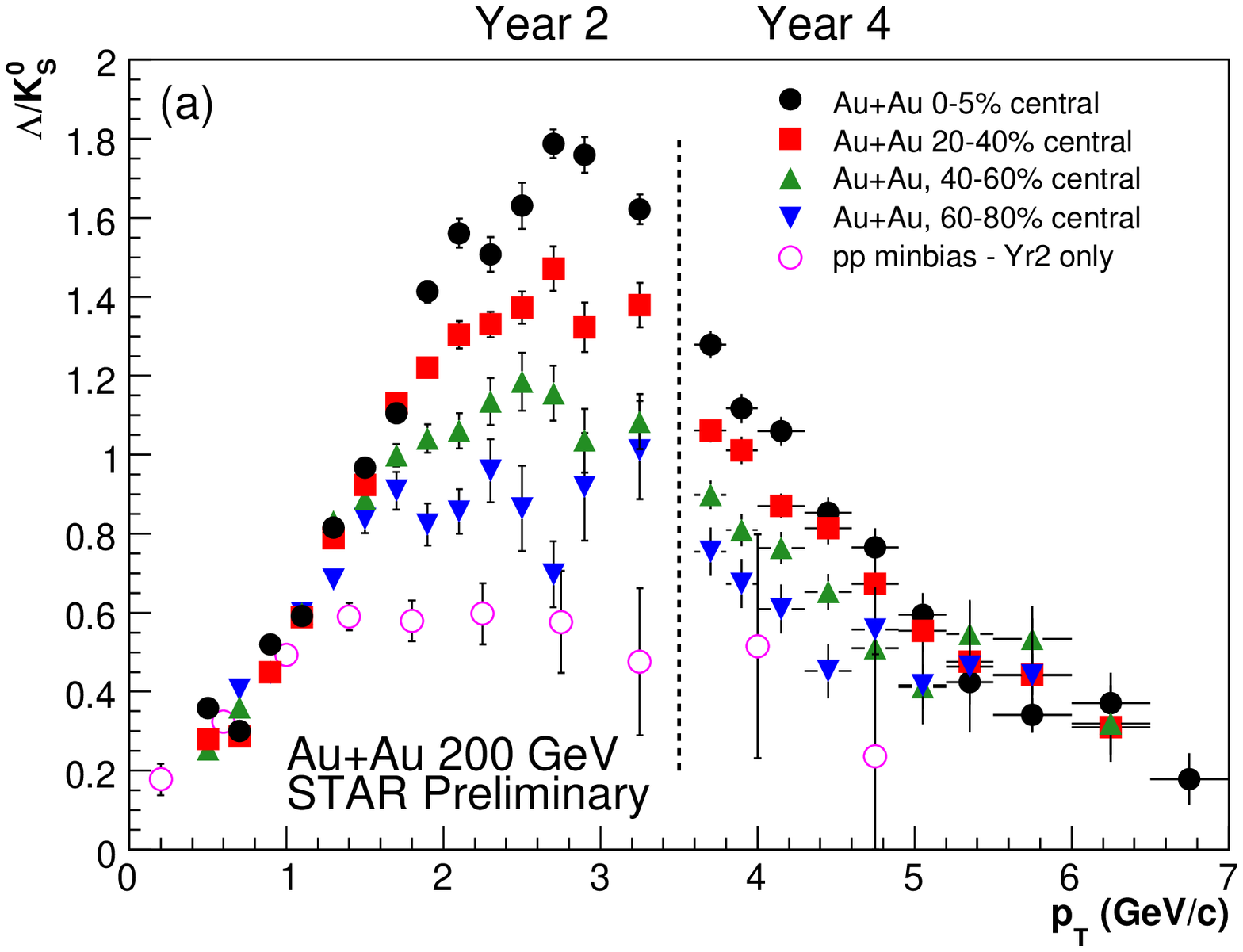}
\caption{$\Lambda/K^{0}_{S}$ ratios measured in p+p and Au+Au collisions~\cite{Bielcikova:2006nv}.}
\label{figure:lambdaoverk}
\end{figure}

\subsection{Particle spectra and ratios}
Majority of the particles emitted in heavy-ion collisions appear in the low
transverse momentum region: $\pt~<$ 2 GeV/c. This region was extensively studied
in AGS and SPS experiments, measurements at higher transverse momenta were limited
by beam energy. At RHIC, the transverse momentum range of identified 
particles is extended to high-$\pt \sim$ 5 GeV/c~\cite{PhenixPbarPi} and 
later up to $\pt \sim$ 12 - 15 GeV/c~\cite{StarPbarPi}.
Experimental results and theoretical calculations suggest distinction of three $\pt$ 
regions: bulk, intermediate and high $\pt$. The bulk region ($\pt$ < 2 GeV/c) seems
to be driven by the thermal properties of the matter created in heavy-ion collisions.
In the high$-\pt$ (6 GeV/c $<\pt$) region particle measured particle spectra
are well described by pQCD calculations. 
Baryon production is significantly enhanced and the baryon/meson ratio reaches 1 in 
most central Au+Au collisions at 200 GeV as shown in Fig.~\ref{figure:pbarpi}.
Expectations of particle production from elementary collisions, quark and gluon jets
and recombination/coalescence mechanisms are also shown~\cite{PhenixPbarPi,StarPbarPi}. 
Similar enhancement is observed in strange particle production 
as shown in Fig.~\ref{figure:lambdaoverk}, however results from p+p collisions
show clear collision energy dependence. 
Coalescence-recombination models quantitatively agree with results
on particle spectra ($R_{AA}$), particle ratios (eg. $\overline{p}/\pi$,
$\Lambda/K^{0}_{S}$) and elliptic flow in the intermediate $\pt$ region. 

\subsection{Azimuthal di-hadron correlations at mid- and forward-rapidities}
\begin{figure}[ht]
\includegraphics[width=0.96\textwidth]{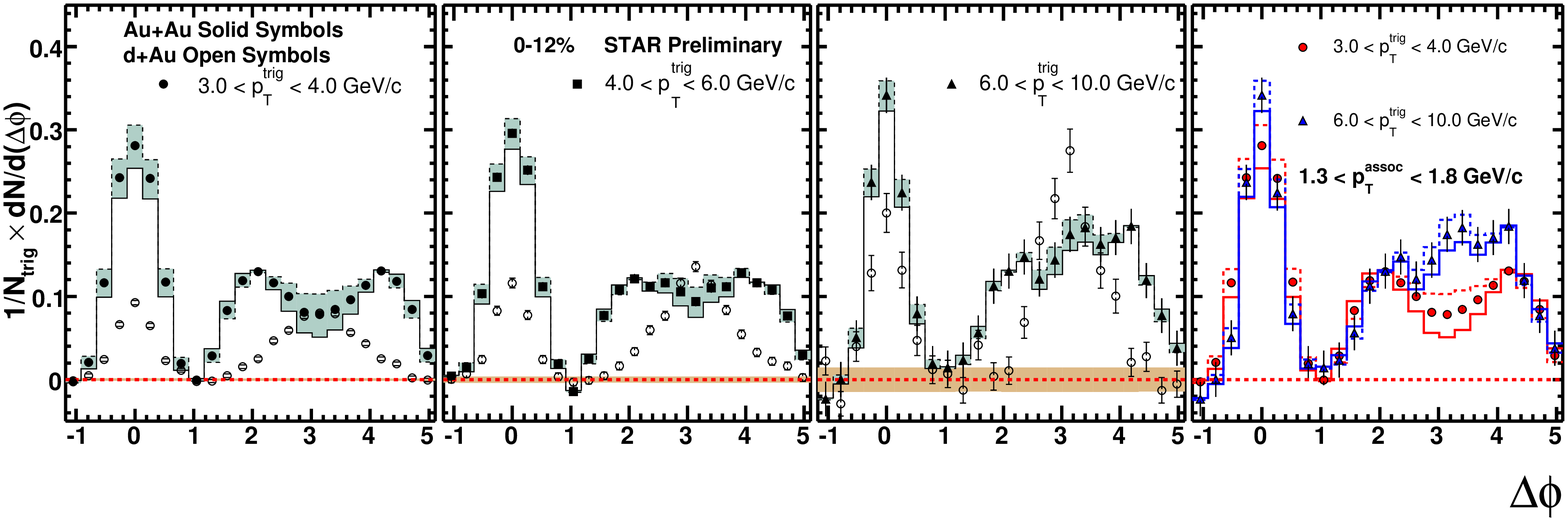}
\caption{Azimuthal di-hadron distributions are shown for different trigger particle selections~\cite{Horner:2007gt}.}
\label{figure:2part}
\end{figure}

While full jet reconstruction only became available recently at RHIC, hard processes 
are studied via high-$\pt$ particle production and their (azimuthal) correlation with 
lower momentum particles. These measurements revealed significant suppression of 
high-$\pt$ hadrons and the disappearance of the away-side jet in central Au+Au 
collisions~\cite{hptsupSTAR,hptsupPHENIX}.
Fig.~\ref{figure:2part} shows a systematic measurement of azimuthal di-hadron correlations
by STAR, where the away-side is significantly broadened with respect to proton-proton 
collisions and the near-side remains unchanged~\cite{Horner:2007gt}. 
The broadening and the modification of the away-side correlation and the presence of 
long range $\Delta\eta$ correlation
on the near-side lead to various theoretical explanations. Models based on the assumption of flow 
deflected jets, large angle gluon radiation~\cite{Vitev:2005yg,Polosa:2006hb}, 
conical flow induced by shock waves in the medium (Mach cones)~\cite{Stoecker:2004qu,CasalderreySolana:2004qm} and 
Cerenkov radiation~\cite{Dremin:2005an,Koch:2005sx} provide alternative descriptions of azimuthal di-hadron correlations.
%
%

%
\begin{figure}[th]   
\begin{center}
\includegraphics[width=0.48\textwidth]{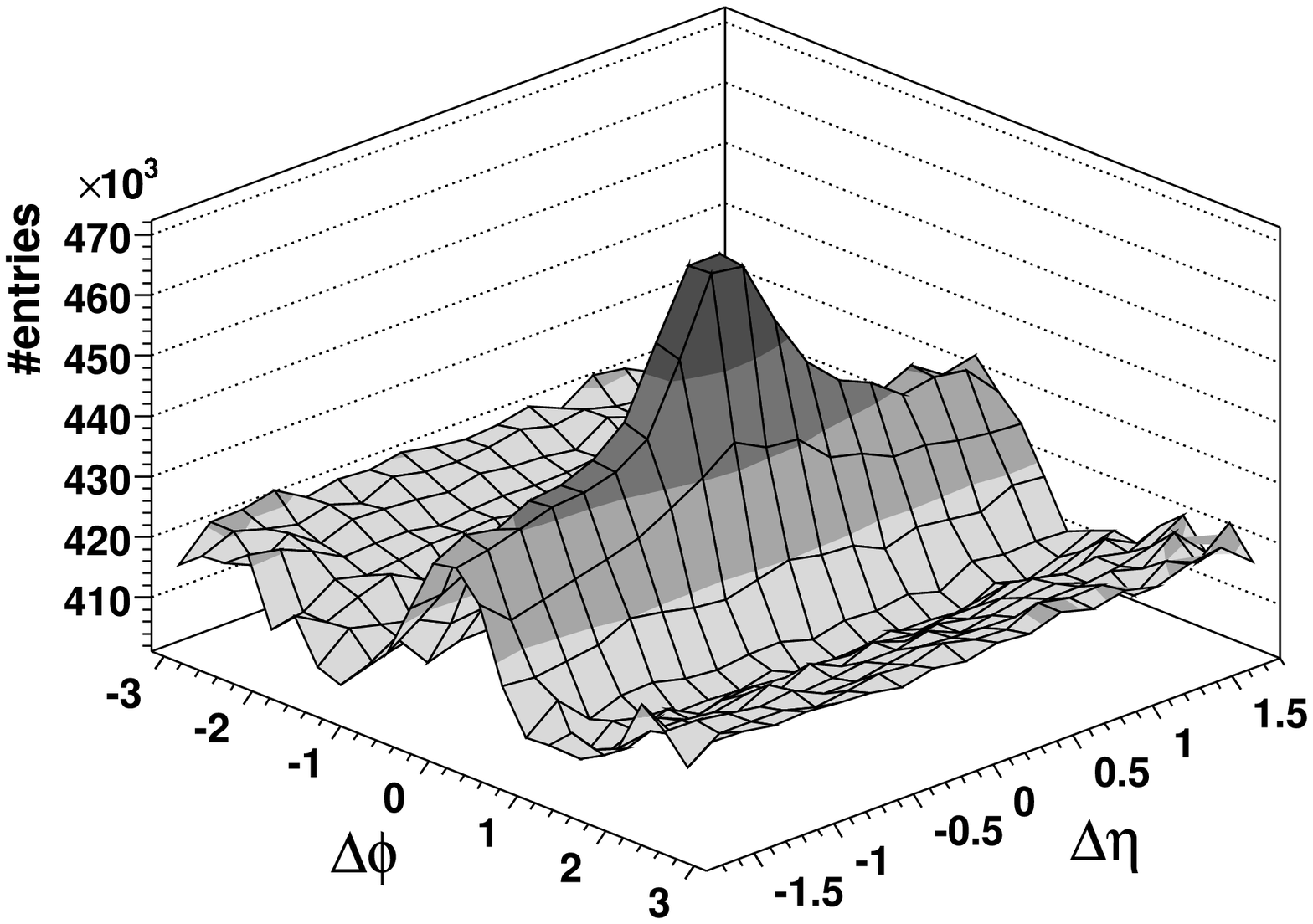}
\includegraphics[width=0.48\textwidth]{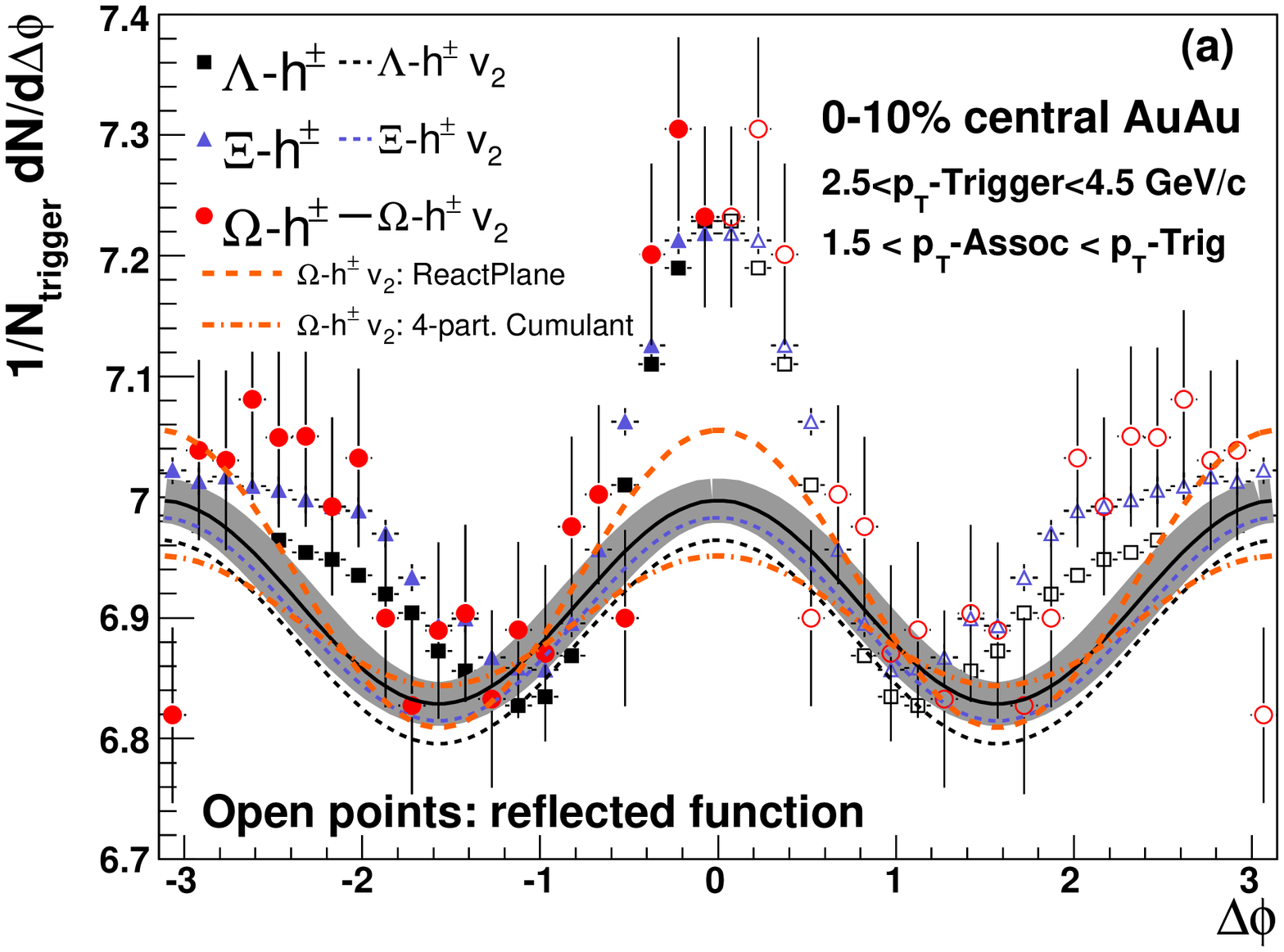}
\caption{Raw $\Delta\eta\times\Delta\phi$ correlation function in Au+Au collisions~\cite{Putschke:2007mi} and strange, multi-strange -  charged hadron correlations in central (0-10\%) Au+Au collisions~\cite{Bielcikova:2007mb}.}
\label{figure:ridge}
\end{center}
\end{figure}
%

%
\begin{figure}[th]   
\begin{center}
\includegraphics[width=0.48\textwidth]{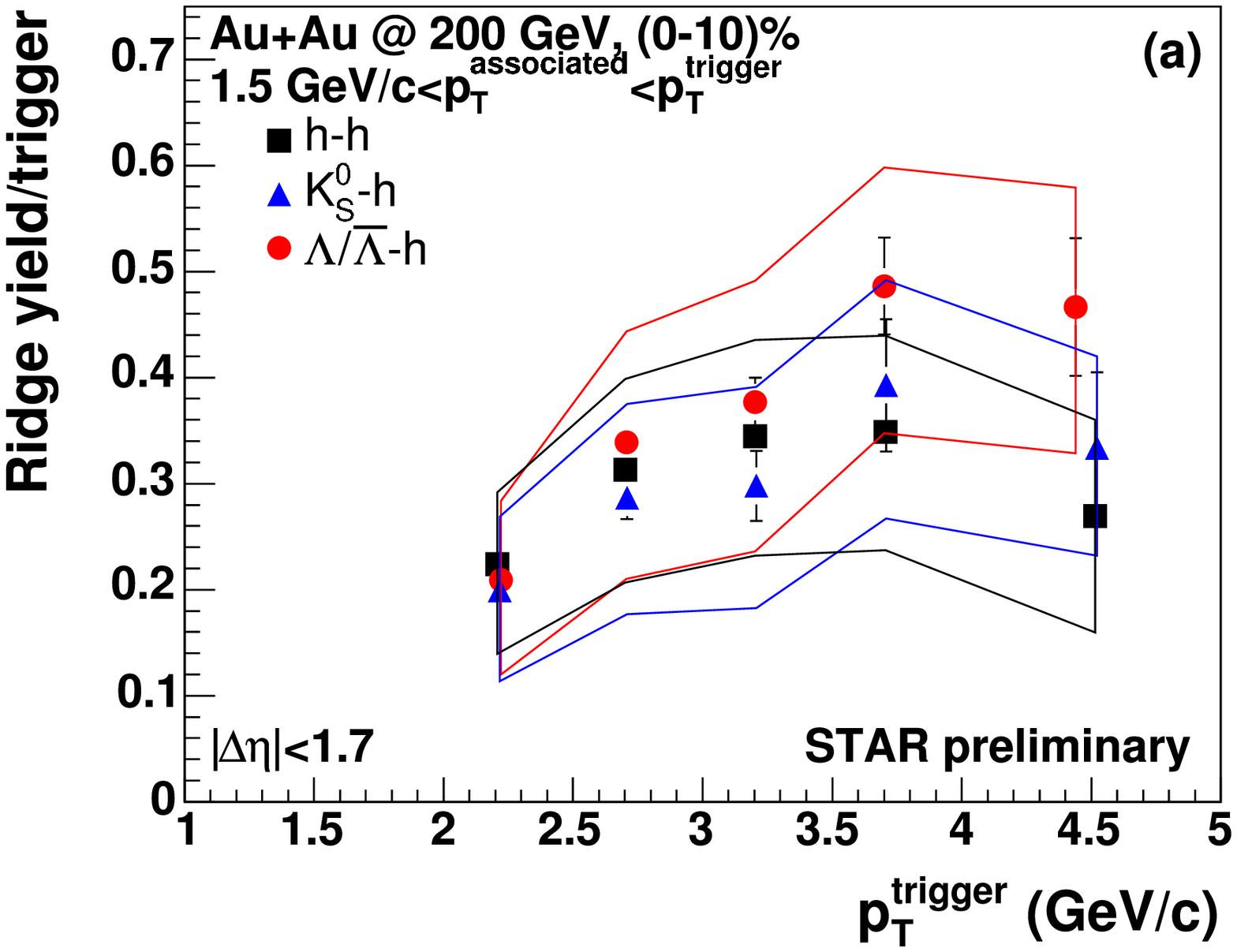}
\includegraphics[width=0.48\textwidth]{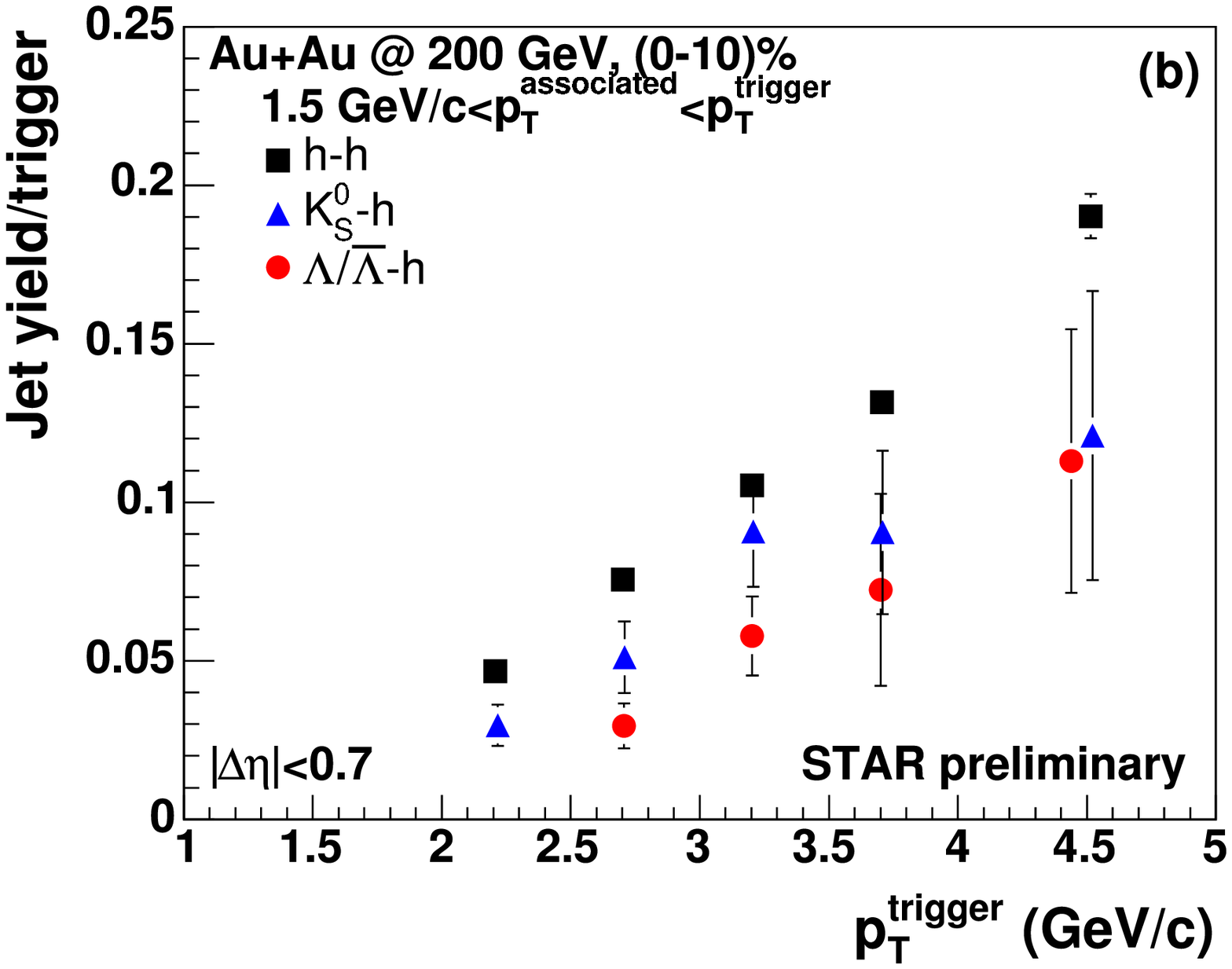}
\caption{Jet and ridge yields as a function of trigger $p_{T}$ in central (0-10\%) Au+Au collisions~\cite{Bielcikova:2007mb}.}
\label{figure:ridgeplusjet}
\end{center}
\end{figure}
%

Extended measurements of the near-side peak in wide $\Delta\eta$
reveal enhancement in the near-side correlated yield extending up to $\left|\Delta\eta\right| \sim$ 
1.5 in Au+Au collisions, as shown in Fig.~\ref{figure:ridge}, which is commonly referred as the ridge~\cite{Putschke:2007mi,Jacobs:2005pk}. 
Two particle correlation measurements at low transverse momenta ($\pt~<$ 2 GeV/c) 
without trigger particle also show long range correlation in 
$\Delta\eta$, above the flow modulated background~\cite{Adams:2004pa}. 
There are many competing theories to describe the long range $\Delta\eta$ correlations: coupling 
of induced radiation to the longitudinal flow~\cite{Armesto:2004pt}, turbulent color 
fields~\cite{Majumder:2006wi}, anisotropic plasma~\cite{Romatschke:2006bb} and the interplay of 
jet-quenching and strong radial flow~\cite{Voloshin:2004th}. 
Recombination of locally thermal enhanced partons due to partonic energy loss also 
provide a ridge like signal~\cite{Chiu:2005ad}. Up to now, STAR has the unique opportunity to fully characterize
the properties of the long range $\Delta\eta$ correlations.
Figure~\ref{figure:ridge} also shows strange and multi-strange hadron triggered azimuthal di-hadron correlations
measured by STAR above the flow background. Clear near-side correlation is shown in contrast to coalescence/recombination
expectations for $\Omega$, however recently a new solution is suggested~\cite{Hwa:2007vp}.
The jet (peak) and the bulk (shoulder structure) is separated and the extracted correlated yields
are shown in Fig.~\ref{figure:ridgeplusjet}. The correlated jet yield is increasing, while the ridge 
yield seems to level off with increasing trigger $p_{T}$~\cite{Bielcikova:2007mb}. The ridge is considered the response of the 
medium to the traversing hard partons/jets. 
The properties of the peak structure are jet like while the shoulder like structure is similar to the bulk.
%

%
\begin{figure}[ht]
\begin{center}
\includegraphics[width=0.5\textwidth]{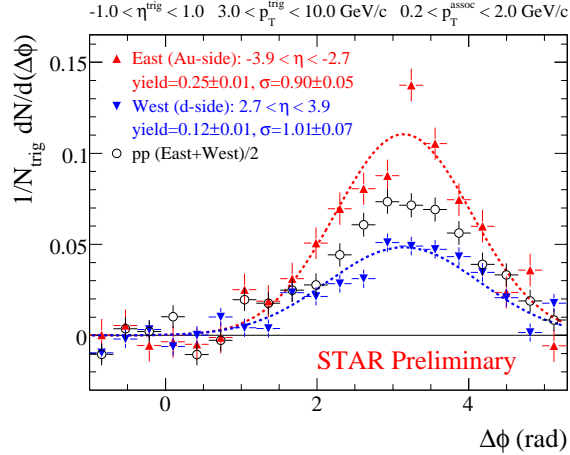}
\caption{Correlation functions for pseudo-rapidity averaged p+p and pseudo-rapidity selected d+Au collision~\cite{Molnar:2007wy}.}
\label{figure:forwppdau}
\end{center}
\end{figure}

STAR has reported measurements in p+p, d+Au and Au+Au collisions 
from the forward rapidity region utilizing its Forward Time Projection 
Chambers~\cite{Ackermann:2002yx} (FTPCs). Extension of the azimuthal di-hadron correlation
measurements to large rapidities have the possibility to explore modifications of away-side 
correlations and the possible existence of long range correlations on the near side by the
bulk response. In these measurements, the high-$\pt$ trigger particles are selected in the STAR Time 
Projection Chamber (TPC): $\left|\Delta\eta\right|$ < 1, while the associated particles
are selected in the FTPCs: 2.7 < $\left|\Delta\eta\right|$ < 3.9. These selections
introduce a minimum 1.7 pseudo-rapidity gap and are possibly sensitive to small-x gluon and large-x
quark hard scattering~\cite{Wang:2006xq}, while azimuthal di-hadron correlations with the selection of the trigger
particle and the associated particles at mid-rapidity address gluon-gluon hard scatterings at 
RHIC energies~\cite{CTEQ}. 
%
%
\begin{figure}[th]   
\begin{center}
\includegraphics[width=0.48\textwidth]{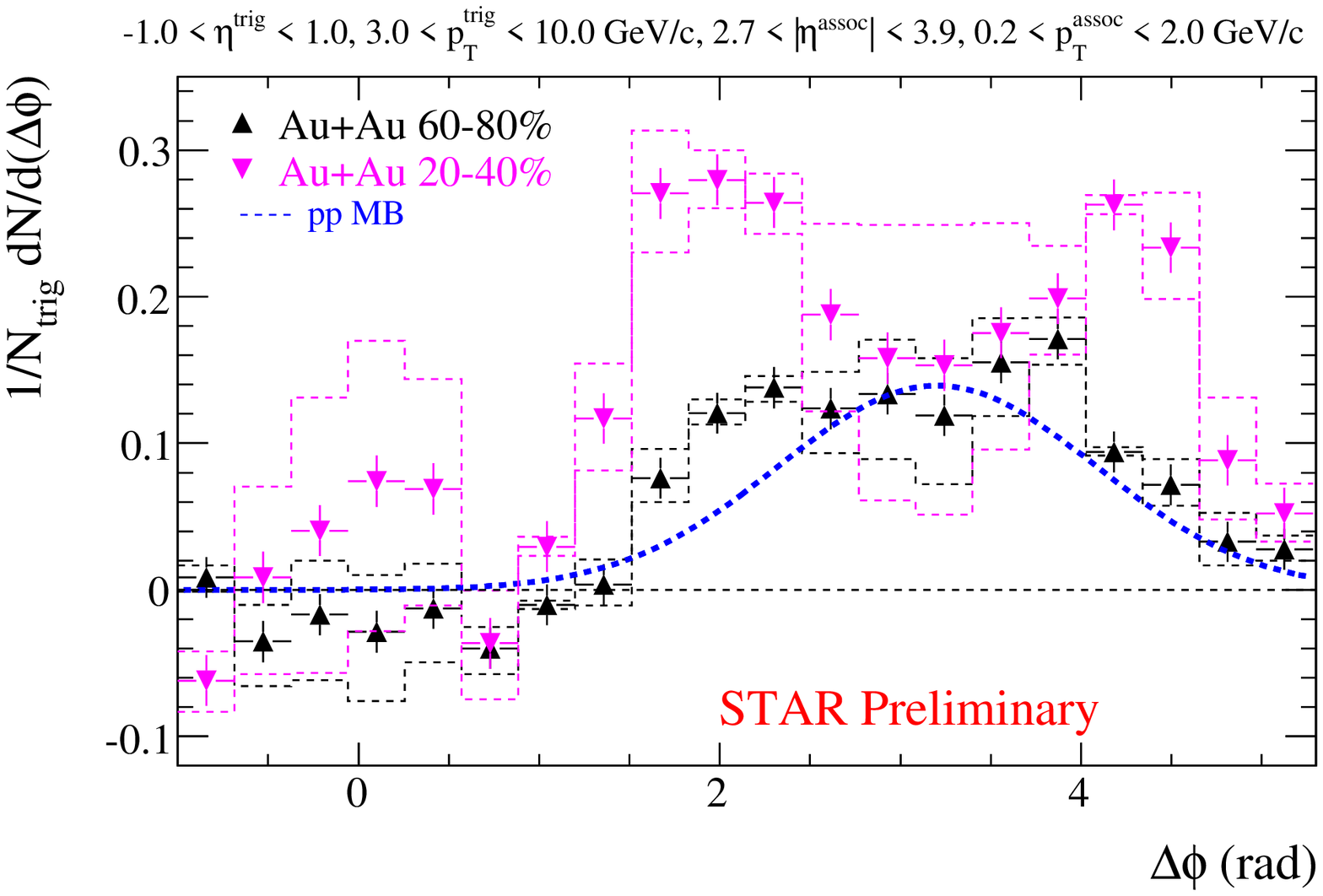}
\includegraphics[width=0.48\textwidth]{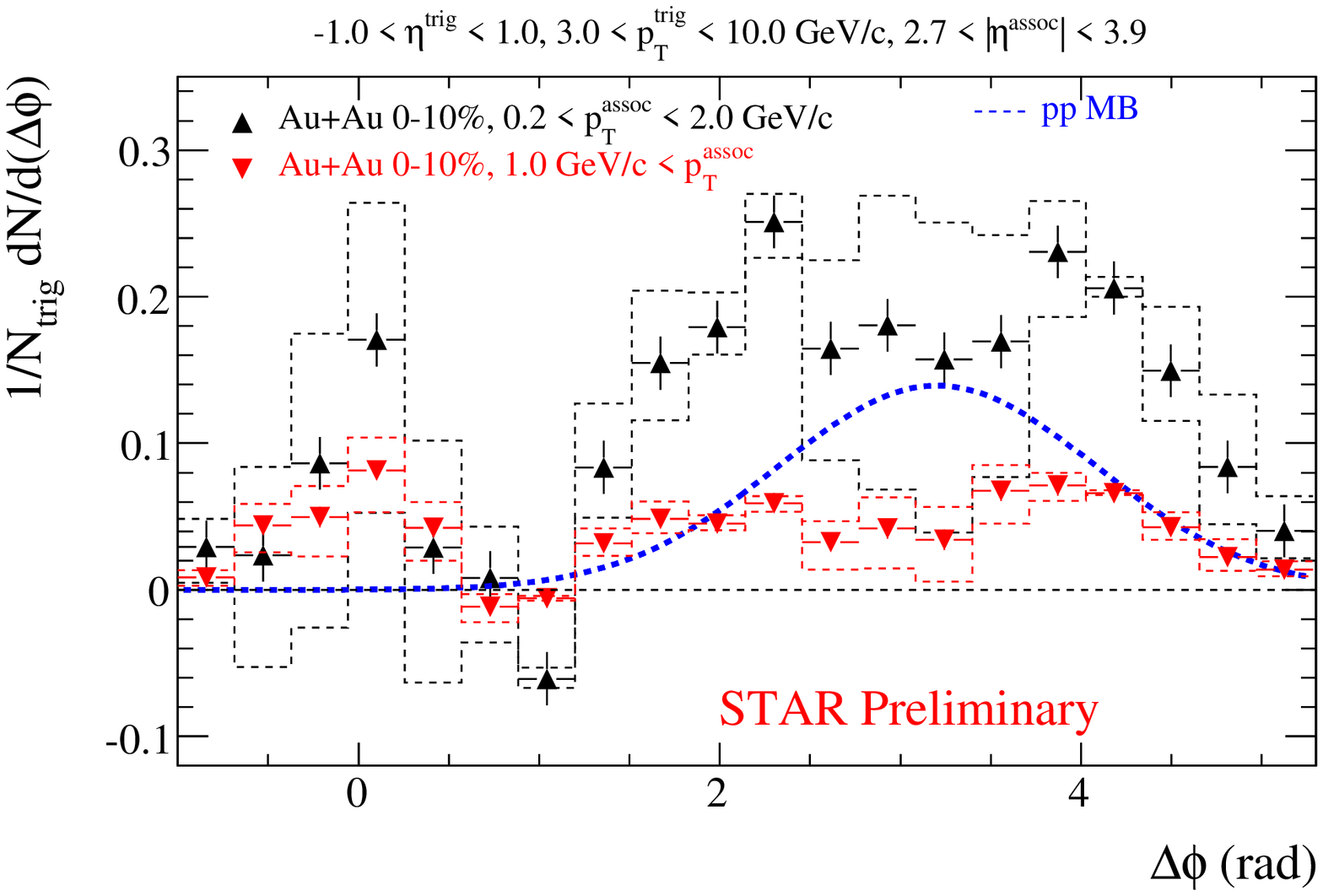}
\caption{Corrected $\Delta\phi$ correlation functions in Au+Au collisions compared to p+p collisions~\cite{Molnar:2007wy}.}
\label{figure:midforAuAu}
\end{center}
\end{figure}
%
Figure~\ref{figure:forwppdau} shows the two particle distributions in p+p and d+Au 
collisions~\cite{Wang:2006xq,Molnar:2007wy} at forward rapidities. In d+Au collisions
the outgoing $d$ and $Au$ sides are shown separately, while the p+p results are averaged over
positive and negative pseudo-rapidities. The near-side correlation is consistent with zero
both for p+p and d+Au collisions. On the away-side a factor of 2 suppression of the d-side yield is 
observed with respect to the Au-side. The p+p results are situated between the d- and Au-side. 
%
%
%
%
\begin{figure}[ht]   
\begin{center}
\resizebox{0.48\textwidth}{!}{\includegraphics{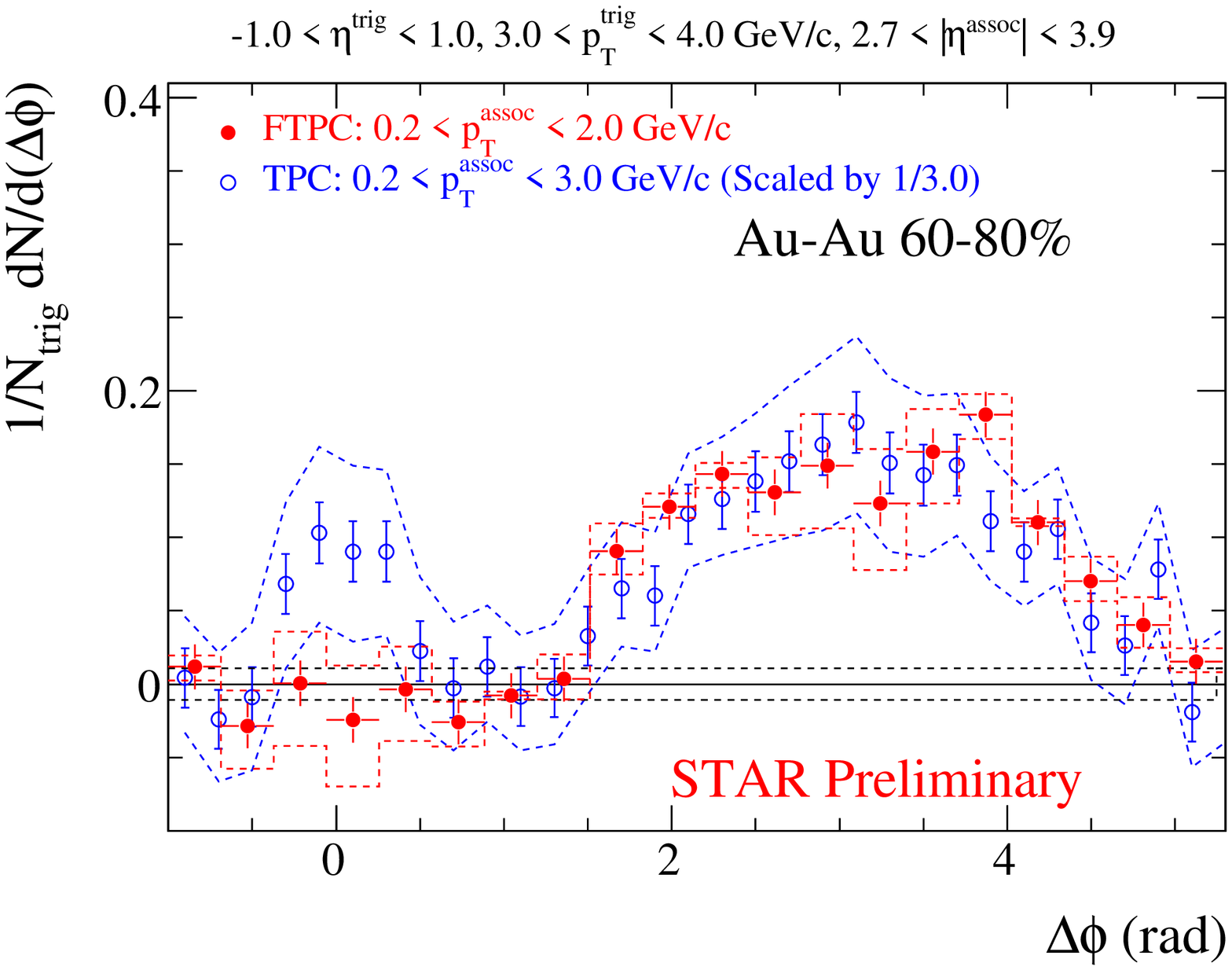}}
\resizebox{0.48\textwidth}{!}{\includegraphics{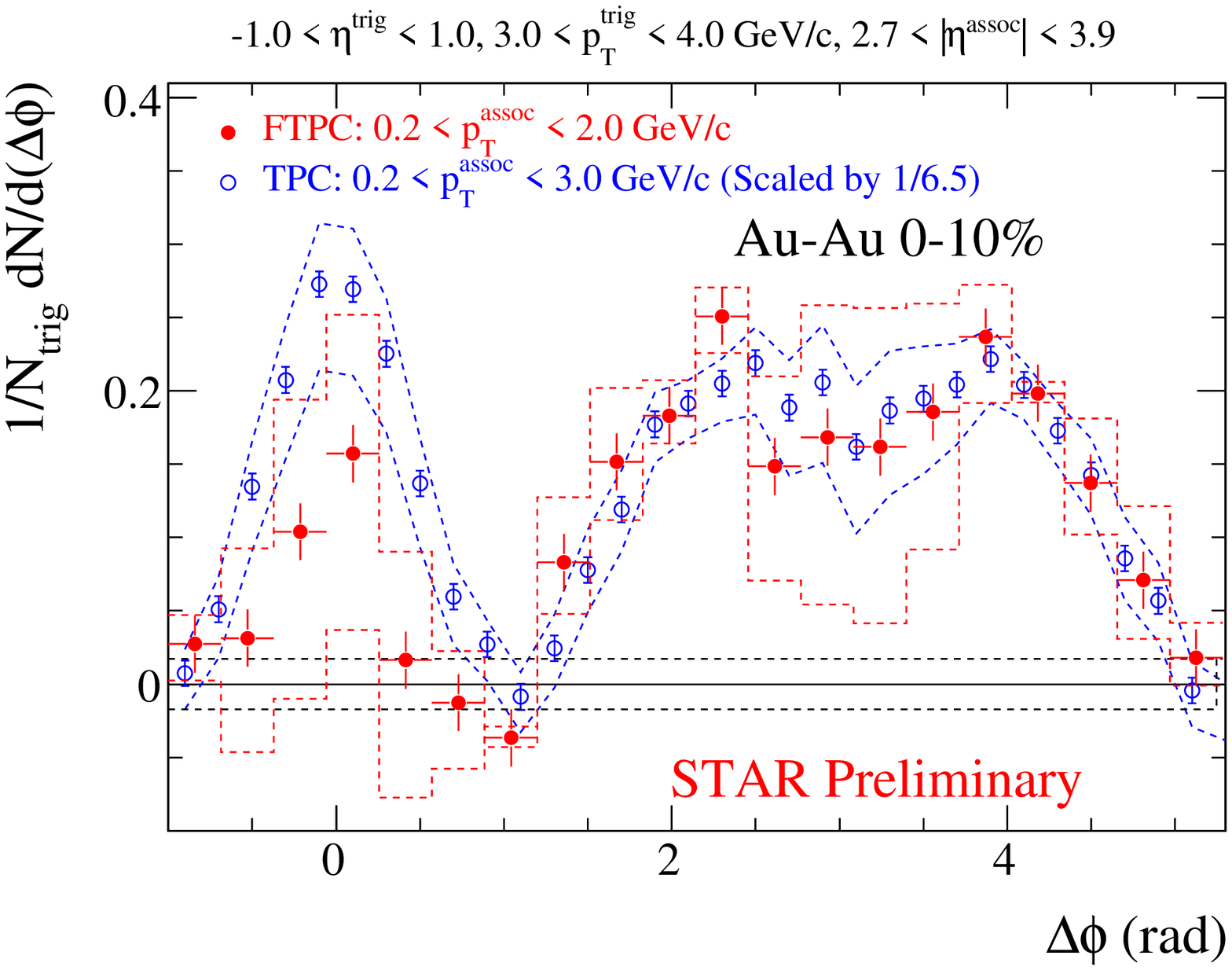}}
\caption{Comparison of azimuthal di-hadron correlation functions from mid- and mid-forward rapidities peripheral and central Au+Au collisions~\cite{Molnar:2007wy}.}
\label{figure:TPCFTPC}
\end{center}
\end{figure}
%
The suppression of the d-side yield is in qualitative agreement with the Color Glass 
Condensate (CGC) picture~\cite{CGC}, which predicts suppression of small-x gluons in 
the Au nucleus. However, the reduction of the d-side yield may arise from the energy 
degradation of the d-side quarks due to multiple scattering in the Au 
nucleus. Gluon anti-shadowing~\cite{antishadowing} and the EMC effect~\cite{EMC}
would enhance the d-side yield relative to the Au-side~\cite{Wang:2006xq,Molnar:2007wy}.
Figure~\ref{figure:midforAuAu} shows the mid- forward-rapidity
azimuthal correlations for different centralities and associated $\pt$ ranges in Au+Au 
collisions at $\s$ = 200 GeV, which can be regarded as the extension of 
Fig.~\ref{figure:ridge} to large pseudo-rapidities. 
In peripheral and mid-peripheral Au+Au collisions the 
near-side correlations for 0.2 $< \pt^{assoc} <$ 2 GeV/c are consistent with zero within
the systematic uncertainties. For the most central (0-10\%) Au+Au dataset two associated
$\pt$ ranges are selected. The near-side correlations for 0.2 $< \pt^{assoc} <$ 2 GeV/c 
are consistent with zero within the systematic uncertainties.
However, the central data at high associated $\pt$, with the reduced systematic uncertainty,
are suggestive of non-zero correlation on the near-side.
This result indicates that long range $\deta$ correlations, first observed in $\deta<$ 2~\cite{Putschke:2007mi}, 
may extend out to $\deta \sim$ 4 in the FTPCs~\cite{Molnar:2007wy}.

In mid- forward-rapidity correlations, broadening of the away-side correlation shapes is observed in 
Au+Au collisions with respect to p+p as shown in Fig.~\ref{figure:midforAuAu}. The broadening is present 
for each centrality, such as at mid-rapidity~\cite{Horner:2007gt}, however the away-side yield at forward-rapidity
is a factor of 3 - 6.5 smaller~\cite{Molnar:2007wy}. Figure~\ref{figure:TPCFTPC} shows the comparison 
of the azimuthal correlations measured at forward rapidity (with the scaled down yield) 
and at mid-rapidity as shown eg. in Fig~\ref{figure:2part}. The away-side correlation shapes measured 
in the STAR-TPC and FTPCs are identical within the systematic uncertainties.  
This might suggests similar apparent energy loss at mid- and forward-rapidities,
however the interplay of the physical processes might be different.

\section{Prospects for LHC}

The LHC will deliver 14 TeV proton-proton collisions and 5.5 TeV Pb+Pb collisions 
at high luminosity. Among the LHC experiments, ALICE has strong capabilities in 
particle identification, down to $\pt \sim$ 200 MeV/c due to the low magnetic 
field (0.2-0.5T) and the material budget.
Particle identification is achieved by the combination of the ALICE sub-detectors:
ITS, TPC, TOF, TRD and HMPID and the more specified EMCAL and PHOS~\cite{Alessandro:2006yt}.

In the first run, LHC will deliver proton-proton collisions, which are not only
baseline for heavy-ion measurements but represent essential physics to contrast 
with theoretical expectations. There are many questions left open in proton-proton
and heavy-ion physics at RHIC for the concise interpretation.
At LHC energies the azimuthal di-hadron correlations will benefit from the enhanced 
cross-section of large-$Q^{2}$ processes, but the uncorrelated background which 
appears in the combinatorial background will also increase. ALICE carried out a feasibility
study on azimuthal di-hadron correlations (leading particle correlations) in simulated 
unquenched and quenched HIJING Pb+Pb events at $\s$ = 5.5 TeV, with a choice of trigger
and associated particle selection to be able to compare to RHIC results~\cite{Ploskon}.  
Trigger bias has to be properly addressed and the $\gamma$ - charged hadron correlation 
provide good solution, however limited in the lower transverse momentum region~\cite{Arleo:2007qw}.
Azimuthal di-hadron correlations in $\Delta\eta$ x $\Delta\phi$ and 3-particle correlations
are not yet address in feasibility studies, which will be important to repeat and extend them
to identified correlations to gain more insights to the interaction of hard scattered partons and
the bulk.

\section{Summary}

Selected review of RHIC results are presented, focusing on particle production
azimuthal di-hadron correlations in the intermediate transverse momentum region.
Two particle correlations in $\Delta\phi$ and $\Delta\eta$ are important tools 
to investigate and characterize the interaction of the hard probes and the medium
created in heavy-ion collisions.
Combined measurement of single particle and correlated yields might help to disentangle
the physics processes and provide solid baseline to contrast with theoretical expectations.


\begin{thebibliography}{99}

\bibitem{hptsupSTAR}
  C.~Adler {\it et al.}  [STAR Collaboration],
  Phys.\ Rev.\ Lett.\  {\bf 89} (2002) 202301 [arXiv:nucl-ex/0206011].
\bibitem{hptsupPHENIX}
  K.~Adcox {\it et al.}  [PHENIX Collaboration],
  Phys.\ Rev.\ Lett.\  {\bf 88} (2002) 022301 [arXiv:nucl-ex/0109003].
\bibitem{AwayDisPhenix}
  C.~Adler {\it et al.}  [STAR Collaboration],
  Phys.\ Rev.\ Lett.\  {\bf 90} (2003) 082302
  [arXiv:nucl-ex/0210033].

\bibitem{AwayDisStar}
  J.~Adams {\it et al.}  [STAR Collaboration],
  Phys.\ Rev.\ Lett.\  {\bf 95} (2005) 152301
  [arXiv:nucl-ex/0501016].

\bibitem{AwayDisCer}
  G.~Agakichiev {\it et al.}  [CERES/NA45 Collaboration],
  Phys.\ Rev.\ Lett.\  {\bf 92} (2004) 032301
  [arXiv:nucl-ex/0303014].

\bibitem{Putschke:2007mi}
  J.~Putschke,
  arXiv:nucl-ex/0701074.
\bibitem{PhenixPbarPi} 
  S.~S.~Adler {\it et al.}  [PHENIX Collaboration],
  Phys.\ Rev.\ Lett.\  {\bf 91} (2003) 172301 [arXiv:nucl-ex/0305036].
\bibitem{StarPbarPi} 
  B.~I.~Abelev {\it et al.}  [STAR Collaboration],
  Phys.\ Rev.\ Lett.\  {\bf 97} (2006) 152301 [arXiv:nucl-ex/0606003].
\bibitem{Bielcikova:2006nv}
  J.~Bielcikova  [STAR Collaboration],
  Nucl.\ Phys.\  A {\bf 783} (2007) 565
  [arXiv:nucl-ex/0612028].

%

\bibitem{Horner:2007gt}
  M.~J.~Horner  [STAR Collaboration],
  arXiv:nucl-ex/0701069.

\bibitem{Vitev:2005yg}
  I.~Vitev,
  Phys.\ Lett.\  B {\bf 630} (2005) 78
  [arXiv:hep-ph/0501255].

%
\bibitem{Polosa:2006hb}
  A.~D.~Polosa and C.~A.~Salgado,
  Phys.\ Rev.\  C {\bf 75} (2007) 041901
  [arXiv:hep-ph/0607295].
  
%
\bibitem{Stoecker:2004qu}
  H.~Stoecker,
  Nucl.\ Phys.\  A {\bf 750} (2005) 121
  [arXiv:nucl-th/0406018].
  
%
\bibitem{CasalderreySolana:2004qm}
  J.~Casalderrey-Solana, E.~V.~Shuryak and D.~Teaney,
  J.\ Phys.\ Conf.\ Ser.\  {\bf 27} (2005) 22
  [Nucl.\ Phys.\  A {\bf 774} (2006) 577]
  [arXiv:hep-ph/0411315].

%
\bibitem{Dremin:2005an}
  I.~M.~Dremin,
  Nucl.\ Phys.\  A {\bf 767} (2006) 233
  [arXiv:hep-ph/0507167].
  
%
\bibitem{Koch:2005sx}
  V.~Koch, A.~Majumder and X.~N.~Wang,
  Phys.\ Rev.\ Lett.\  {\bf 96} (2006) 172302
  [arXiv:nucl-th/0507063].

\bibitem{Jacobs:2005pk}
  P.~Jacobs,
  Eur.\ Phys.\ J.\  C {\bf 43} (2005) 467
  [arXiv:nucl-ex/0503022].

\bibitem{Adams:2004pa}
  J.~Adams {\it et al.}  [STAR Collaboration],
  Phys.\ Rev.\  C {\bf 73} (2006) 064907
  [arXiv:nucl-ex/0411003].
  
\bibitem{Armesto:2004pt}
  N.~Armesto, C.~A.~Salgado and U.~A.~Wiedemann,
  Phys.\ Rev.\ Lett.\  {\bf 93} (2004) 242301
  [arXiv:hep-ph/0405301].

\bibitem{Majumder:2006wi}
  A.~Majumder, B.~Muller and S.~A.~Bass,
  arXiv:hep-ph/0611135.
  
%
\bibitem{Romatschke:2006bb}
  P.~Romatschke,
  Phys.\ Rev.\  C {\bf 75} (2007) 014901
  [arXiv:hep-ph/0607327].

\bibitem{Voloshin:2004th}
  S.~A.~Voloshin,
  Nucl.\ Phys.\  A {\bf 749} (2005) 287
  [arXiv:nucl-th/0410024].

\bibitem{Chiu:2005ad}
  C.~B.~Chiu and R.~C.~Hwa,
  Phys.\ Rev.\  C {\bf 72} (2005) 034903
  [arXiv:nucl-th/0505014].

\bibitem{Hwa:2007vp}
  R.~C.~Hwa,
  arXiv:nucl-th/0701018.

\bibitem{Bielcikova:2007mb}
  J.~Bielcikova,
  arXiv:nucl-ex/0701047.
\bibitem{Ackermann:2002yx}
  K.~H.~Ackermann {\it et al.},
  Nucl.\ Instrum.\ Meth.\  A {\bf 499} (2003) 713
  [arXiv:nucl-ex/0211014].
\bibitem{Wang:2006xq}
  F.~Wang,
  Nucl.\ Phys.\  A {\bf 783} (2007) 157
  [arXiv:nucl-ex/0610011].


\bibitem{CTEQ}
H.~L.~Lai {\it et al.}  [CTEQ Collaboration],
  Eur.\ Phys.\ J.\ C {\bf 12}, 375 (2000).
  [arXiv:hep-ph/9903282].

\bibitem{Molnar:2007wy}
  L.~Molnar,
  J.\ Phys.\ G {\bf 34} (2007) S593
  [arXiv:nucl-ex/0701061].
\bibitem{CGC}
 L.~V.~Gribov, E.~M.~Levin and M.~G.~Ryskin,
  Phys.\ Rept.\  {\bf 100}, 1 (1983);
  D.~Kharzeev, E.~Levin and L.~McLerran,
  Nucl.\ Phys.\ A {\bf 748}, 627 (2005).
  [arXiv:hep-ph/0403271].


\bibitem{antishadowing}
  K.~J.~Eskola, V.~J.~Kolhinen and P.~V.~Ruuskanen,
  Nucl.\ Phys.\ B {\bf 535}, 351 (1998).
  [arXiv:hep-ph/9802350].
  
\bibitem{EMC}
 J.~J.~Aubert {\it et al.}  [European Muon Collaboration],
  Phys.\ Lett.\ B {\bf 123}, 275 (1983);
   J.~Gomez {\it et al.},
  Phys.\ Rev.\ D {\bf 49}, 4348 (1994).


\bibitem{Alessandro:2006yt}
  B.~Alessandro {\it et al.}  [ALICE Collaboration],
  J.\ Phys.\ G {\bf 32} (2006) 1295.

\bibitem{Ploskon}
H.~Appelshauser~and~M.~Ploskon, ALICE-INT-2005-49.



\bibitem{Arleo:2007qw}
  F.~Arleo,
  arXiv:hep-ph/0701207.
  

\end{thebibliography}
\end{document}